\documentclass{aa}
\usepackage{graphicx,graphics,natbib}
\usepackage{txfonts}
\begin{document}

   \title{Long-term variation in distribution of sunspot groups}

   \author{E. Forg\'acs-Dajka\inst{1,2}
          \and
          B. Major\inst{1}
	  \and
	  T. Borkovits\inst{3}}

   \offprints{E. Forg\'acs-Dajka}

   \institute{E\"otv\"os University, Department of Astronomy , 
              H-1518 Budapest, Pf. 32., Hungary\\
              \email{E.Forgacs-Dajka@astro.elte.hu, B.Major@astro.elte.hu}
         \and
              Konkoly Observatory of HAS, 
              H-1525 Budapest, Pf. 67, Hungary
	  \and
	      Baja Astronomical Observatory of B\'acs-Kiskun County,
	      H-6500 Baja, Szegedi \'ut, Kt. 766, Hungary\\
	      \email{borko@alcyone.bajaobs.hu}}

   \date{Astronomy and Astrophysics, v.424, p.311-315 (2004)}

   \abstract{
   We studied the relation between the distribution of sunspot groups and the 
   Gleissberg cycle. As the magnetic field is related to the area of the sunspot groups, we used 
   area-weighted sunspot group data. On the one hand, we confirm the previously reported long-term cyclic 
   behaviour of the sum of the northern and southern sunspot group mean latitudes, although we 
   found a somewhat longer period 
   ($P\sim 104$ years). We introduced the difference between the ensemble average area of sunspot groups 
   for the two hemispheres, 
   which turns out to show similar behaviour. We also investigated a further aspect of the Gleissberg 
   cycle where while in the $19$th
   century the consecutive Schwabe cycles are sharply separated from each other, one century later 
   the cycles overlap each 
   other more and more. 
   }

   \maketitle
%

\section{Introduction}

The activity of the Sun has been studied in several ways for a long time. The oldest observed and recorded features on the solar disk are the sunspots. Regular sunspot observations were started by Galileo in 1610 and since then the data sets have been being continuously extended. Using these databases one has the opportunity to study the statistical properties of solar activity \citep[see e.g.][]{vitinsky_etal86}. 
The most prominent cycle in the sunspot series and in all solar activity is the 11-year Schwabe cycle. The long-term trend in the amplitude of the Schwabe cycle is known as the secular Gleissberg cycle \citep{gleissberg45}, which has a varying time scale of 80-120 years. 

This paper aims to study the relation between the distribution of sunspot groups and the Gleissberg cycle. Although the origin of the Gleissberg cycle is unclear, it is generally accepted that the interaction of the solar differential rotation and the magnetic field plays a basic role in the generation of all solar activity. From the observational data sets a correlation can be seen between secular variations in magnetic activity and differential rotation variations \citep{vitinsky_etal86,Yoshimura_Kambry93,Javaraiah03}. Although the role of the differential rotation in the long-term variation is not yet clear, several authors have investigated the interplay of differential rotation and the large-scale magnetic field, which drives the solar grand minima. Different modulational mechanisms have been proposed in the recent literature, namely (1) fluctuations in the $\alpha$-effect which could cause the dynamo to stochastically flip and lead to modulation of magnetic energy; (2) a dynamic $\alpha$-effect forced modulation; (3) two-layer dynamos with meridional circulation; (4) nonlinear dynamics via macroscopic (e.g.~Lorentz force) or microscopic (e.g.~$\Lambda$-quenching) feedback \citep[see a detailed review in][]{tobias02}. 

The following study may provide some additional observational background for the investigation of solar dynamo models dealing with the description of the long-term evolution of solar activity. 


\section{Data and method}

We need a detailed and homogenous sunspot catalogue to examine the fine structure of the long-term variation of the sunspot group distribution, so we use the sunspot data of the Greenwich Photoheliographic Results (henceforth GPR) such as the area, the heliographic position and the time of observation. 

\begin{figure}
\centering
\includegraphics[width=0.9\linewidth]{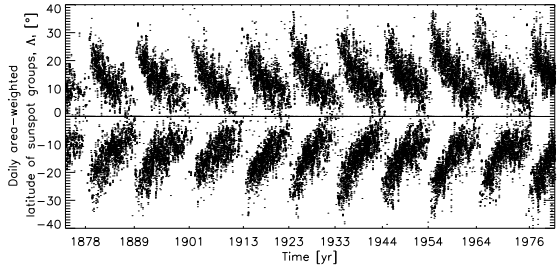}
\caption{The daily area-weighted latitude of sunspot groups as a function of time.}
\label{fig:cshfull_intime}
\end{figure}

\begin{figure}
\centering
\includegraphics[width=0.9\linewidth]{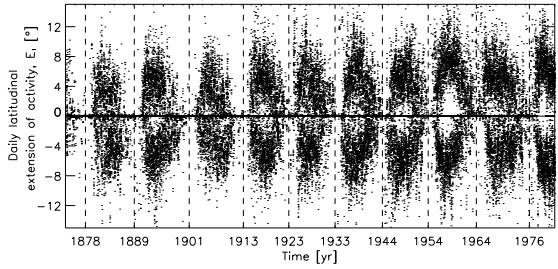}
\caption{The daily latitudinal area-weighted extension of activity as a function of time. The vertical dashed lines denote the Schwabe minima.}
\label{fig:avszfull_intime}
\end{figure}

First, we would like to reanalyse the long term cyclic behaviour of the (signed) sum of the mean latitudes of sunspot groups in the northern and southern hemispheres, respectively, 
but we weighted the latitudinal distribution of sunspot groups with their area.
In a previous work, \citet{pulkkinenetal99} considered 10-day averages of these sums and they found that the ``magnetic equator'' of the Sun displays a long-term variation between northern and southern sunspot group latitudes with a period of $P=33900\pm950$ days, and an amplitude of $1.31\pm0.13$ degrees. They obtained similar results with the yearly averages. 
An important difference with respect to the work of \citet{pulkkinenetal99} is that, we use only Greenwich data. Consequently, our data series is shorter (108 years) than of that of \citet{pulkkinenetal99} (143 years), but is homogeneous, so we can avoid the question of the compatibility of the measurements.

It is well known that solar activity is correlated with the toroidal magnetix flux generated by the solar dynamo; basically the larger the area of sunspot groups, the greater the toroidal magnetic flux. Hence, in order to get a more realistic picture of the distribution of the toroidal magnetic flux, it is better to consider the latitudinal distribution of sunspot groups weighted by their area. For this reason, we define the following expression as the area-weighted mean latitude of sunspot groups:
\begin{equation}
\langle\Lambda\rangle_t=\frac{\sum_{i=1}^{N_t} A_i \Phi_i}{\sum_{i=1}^{N_t} A_i},
\end{equation}
where $A_i$ refers to the area of $i$th sunspot group, $\Phi_i$ denotes the latitude of $i$th sunspot group, $N_t$ is the number of sunspot groups in the examined time intervals in day, and $t$ refers to the length of the examined time interval. We consider 2-day, monthly and yearly (i.e.~$t=2$, $t=30$, $t=365$, respectively) area-weighted mean latitude data series for both the northern and southern hemispheres. On the one hand, we have not found significant differences between the results obtained with different time intervals data sets. On the other hand, the average lifetime of sunspot groups is about one month, so we concentrate only on the 30-day area-weighted averaged data.

Secondly, we would like to investigate the correlation between the Gleissberg cycle and the distribution of sunspot groups. In this case, we do not compose the time average, i.e.~$t=1$. The daily area-weighted latitude of sunspot groups as a function of time is shown in Fig.~\ref{fig:cshfull_intime}. This figure indicates that near the different Schwabe minima the distributions of sunspot groups deviate from each other. To analyse this discrepancy, we determine the daily latitudinal area-weighted extension of activity, which can be written in the following form:
\begin{equation}
E_t=\sqrt{\frac{\sum_{i=1}^{N_t} (\langle\Lambda\rangle_t-\Phi_i)^2}{N_t}}.
\label{eq:E}
\end{equation}
Fig.~\ref{fig:avszfull_intime} illustrates the daily latitudinal area-weighted extension of activity as a function of time. In this case the discrepancy between the Schwabe minima is more significant, suggesting that as a Gleissberg cycle tends to its maximum, the consecutive cycles overlap each other more and more, which will be discussed in the next section.


\section{Results and discussion}

\subsection{Long-term behaviour of area-weighted mean latitudes of sunspot groups}

As it was mentioned in the previous section, we reanalysed the long-term variation of the (signed) sum of the mean latitudes of sunspot groups in the northern and southern hemispheres. An important difference to the work of \citet{pulkkinenetal99} is that we considered the latitudinal distribution of sunspot groups weighted by their area. As a first step of our analysis we applied the discrete Fourier-transform (DFT) method for the data. 
The main peak belongs to the period of $P_\mathrm{DFT}=42\,340\pm1\,000$ days. This is significantly larger than the period of \citet{pulkkinenetal99}, even larger than the length of our $39\,000$ day-long observing window. 
As in such a case the fundamental frequency in the power spectrum is misplaced by aliases, we also searched for the period by phase-dispersion (PHD) method.
This resulted in a period of $P_\mathrm{PHD}=37\,900\pm100$ days. The difference between the two frequencies is  $|\nu_\mathrm{DFT}-\nu_\mathrm{PHD}|=3\times10^{-6}$ day$^{-1}$. The reason can be understood from the properties of the spectral window of the data. The fundametal frequency of the DFT is $\nu_\mathrm{DFT}=2.3618\times10^{-5}$ day$^{-1}$.  The full width at half magnitude of the peak, which depends on the length of the data series, is about $2\times10^{-5}$ day$^{-1}$. Consequently, the frequency found by the phase-dispersion method is located very close to the peak.
Furthermore, as it is well known, if the spectral window is symmetrical to the origin, a peak presents also at around $-\nu_\mathrm{DFT}$. This peak, as well as its side lobes can modify the exact location of the fundamental frequency in the spectrum. 

\begin{figure}[t]
\centering
\includegraphics[width=0.7\linewidth]{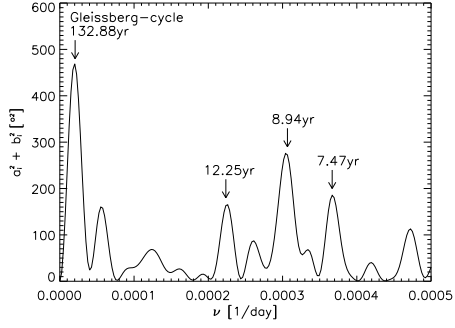}
\caption{The DFT-based power spectrum for the difference between the ensemble average area of sunspot groups for the two hemispheres, i.e.~$(A_{30}/N_{30})_{\mathrm{North}}-(A_{30}/N_{30})_{\mathrm{South}}$.}
\label{fig:dft_hullam_a_sn}
\end{figure}

\begin{figure}
\centering
\includegraphics[width=0.9\linewidth]{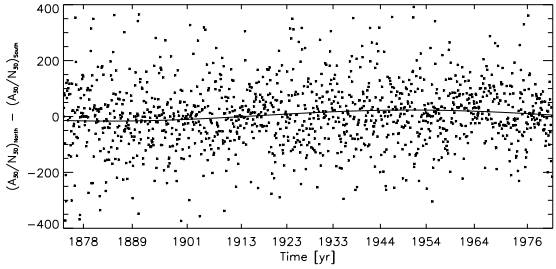}
\caption{Difference between the ensemble average area of sunspot groups for the two hemispheres, i.e.~$(A_{30}/N_{30})_{\mathrm{North}}-(A_{30}/N_{30})_{\mathrm{South}}$. Each dot denotes a value for a 30-day interval, and the solid line represents the fitted sinusoidal profile prescribed in Eq.~\ref{eq:f_t}, with the following parameter values: $\nu=1/48500$, $a_0=3.283199$, $a_1=-6.971889$ and $b_1=-18.475170$.}
\label{fig:hullam_a_sn}
\end{figure}

\begin{table}
\caption[]{The parameters of the sinusoidal curve fitted to the sum of the mean latitudes of sunspots of the two hemispheres by linear least squares for three different periods. Numbers in parenthesis denote the formal errors in the last digits.}
\label{tab:Fourier_lat}
$$
\begin{array}{llll}
\hline
\noalign{\smallskip}
&\mathrm{PHD}&\mathrm{DFT}&\mbox{\citet{pulkkinenetal99}}\\
\noalign{\smallskip}
\hline
\noalign{\smallskip}
P\mbox{ [d]}&37\,900\pm100&42\,340\pm1\,000&33\,900\pm950\,^{\mathrm{a}}\\
a_0\mbox{ [}\degr\mathrm]&-0.331706(28)&-0.383840(28)&-0.303106(29)\\
a_1\mbox{ [}\degr\mathrm]&1.013833(39)&0.337501(40)&1.414234(39)\\
b_1\mbox{ [}\degr\mathrm]&-1.043614(39)&-1.323759(39)&-0.361780(40)\\
\chi^2&6.3164324&6.3197944&6.3222383\\
\noalign{\smallskip}
\hline
\noalign{\smallskip}
\end{array}
$$
\begin{list}{}{}
\item[$^{\mathrm{a}}$] Note: For fitting we use only the period, because their fitted sinusoidal profile did not contain a constant parameter \citep[see Eq.~1 in][]{pulkkinenetal99}
\end{list}
\end{table}

Supposing that the mathematical form of the sinusoidal variation is the following:
\begin{equation}
f(t)=a_0+a_1\cos(2\pi\nu t)+b_1\sin(2\pi\nu t),
\label{eq:f_t}
\end{equation}
we list in Table \ref{tab:Fourier_lat} the values of the Fourier coefficients obtained by least-squares fits, using the previously found periods, as well as the period given in the paper of \citet{pulkkinenetal99}, and the $\chi^2$ values obtained by $\chi^2$ test.

\subsection{Long-term behaviour of the ensemble average area of sunspot groups}

To confirm previous results we would like to investigate the long-term variation of the ensemble average area of sunspot groups. As we mentioned, the toroidal magnetic flux correlates to the sunspot area, so we construct another dataset in the following way. We take the total area ($A_{30}$) and the number of sunspot groups ($N_{30}$) in every 30 days for the northern and southern hemispheres, respectively. We derived the difference between the ensemble average area of sunspot groups for the two hemispheres, as  $(A_{30}/N_{30})_{\mathrm{North}}-(A_{30}/N_{30})_{\mathrm{South}}$. Applying the DFT and the phase dispersion method for these data, the same fundamental period was found (within the error range), namely $P=48\,500\pm100$ days.
The values of the Fourier-coefficients obtained by least-squares fits are $a_0=3.283199(29)$, $a_1=-6.971889(46)$ and $b_1=-18.475170(36)$, while the $\chi^2$ value is $\chi^2=130.0788751$. 

Our results suggest that the toroidal magnetic flux shows a long-term variation between the northern and southern hemispheres, that is, if the ensemble average area of sunspot groups is larger on the southern hemisphere than on the northern one, then we can assume that the toroidal magnetic flux is also larger in the south, and later this dominance alternates.

\subsection{A further aspect of the Gleissberg cycle}

In Figs.~\ref{fig:cshfull_intime}-\ref{fig:avszfull_intime}, it can be seen that the consecutive cycles overlap more and more around the Schwabe minima. In order to study the overlapping areas of the consecutive cycles on a long time scale, we investigated the behaviour of $E_t$ (see Eq.~\ref{eq:E}) around the minima of 11-year cycles. Thus, we considered two-year intervals centered on the individual minima defined by the relative sunspot number \citep[see review in][]{usoskin03}. Then we calculated the mean and the standard deviation of the daily latitudinal area-weighted extension of activity ($\langle E\rangle_{\mathrm{min}}$ and $\sigma_{E,\mathrm{min}}$, respectively) for these intervals.

For our study we invoke the results of \citet{Garcia_Mouradian98}. They calculated the secularly smoothed sunspot number of maxima $R_M^{\ast}(N)$ and minima $R_m^{\ast}(N)$, where $N$ is the 11-yr solar cycle number and they studied the variation of these computed values in time from the viewpoint of the Gleissberg cycle. Accordingly, their results are well comparable with ours.

\begin{figure*}[t]
\centering
\includegraphics[width=0.35\linewidth]{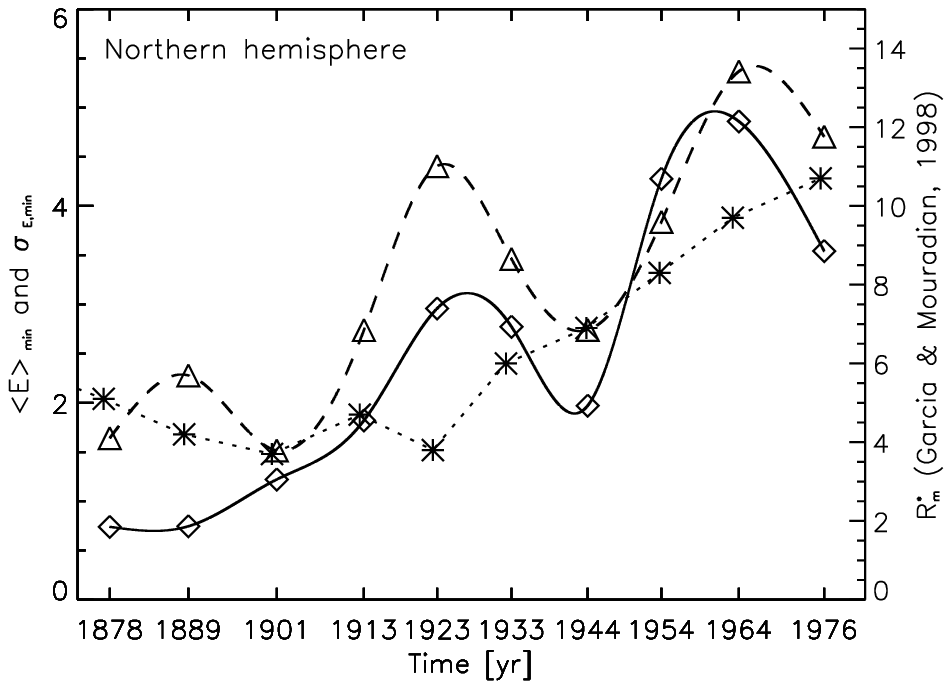}
\hspace*{2cm}
\includegraphics[width=0.35\linewidth]{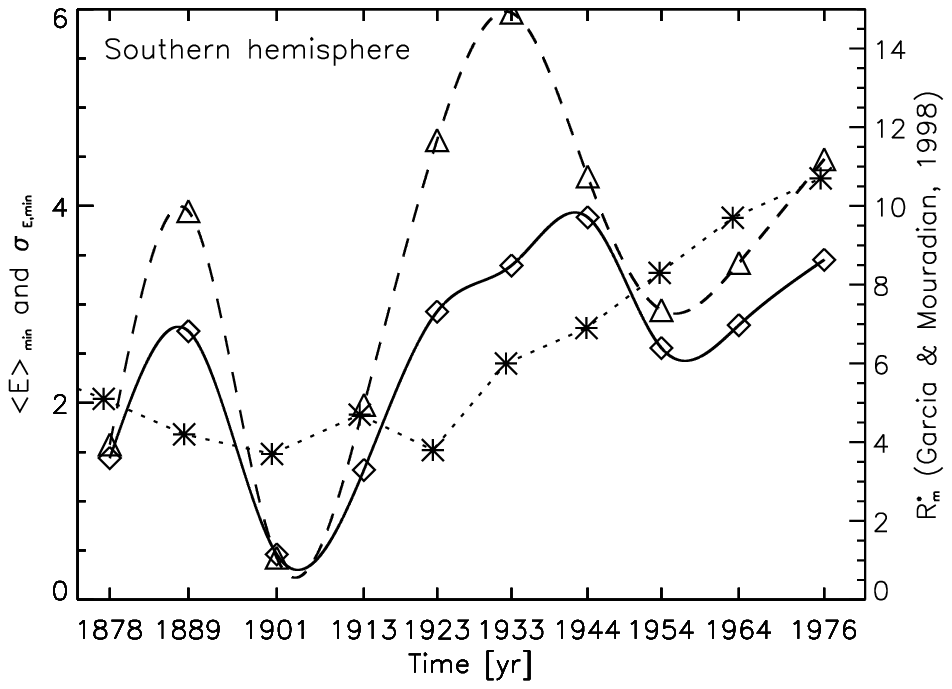}
\caption{The mean (solid line) and the standard deviation (dashed line) of the daily latitudinal extension of activity as a function of time. The dotted line represents the results of \citet{Garcia_Mouradian98}, i.e.~the secularly smoothed sunspot number of minima $R_m^{\ast}$.}
\label{fig:avsz_mean_stdev}
\end{figure*}

In Fig.~\ref{fig:avsz_mean_stdev}, we present the mean ($\langle E\rangle_{\mathrm{min}}$) and the standard deviation ($\sigma_{E,\mathrm{min}}$) of the daily latitudinal area-weighted extension of activity around the minima of 11-yr solar cycle  and the results of \citet{Garcia_Mouradian98}, namely $R_m^{\ast}(N)$ 
It is well visible that as a Gleissberg cycle tends to its maximum $\langle E\rangle_{\mathrm{min}}$ increases. While in the northern hemisphere this tendency is obvious, in the southern one it not so, but the tendency is visible. 
Both these numbers show that the consecutive cycles overlap each other more and more as the Gleissberg cycle approaches its maximum, i.e.~the sunspot groups of the next cycle appear before the end of the previous cycle.


\section{Conclusion}

In this paper we studied the long-term evolution of the distribution of sunspot groups from different aspects, e.g.~latitude, area and extension of activity. We applied two new methods, which in our opinion give a more realistic picture of the long-term variation of the toroidal magnetic field. Our method is based on the results of \citet{Petrovay_Szakaly99}. They found that the latitudinal distribution of the magnetic field at the surface reflects the conditions at the bottom of the convective zone, i.e.~in this regard the convective zone behaves as a ``steamy window''. Although their original statement is restricted to the poloidal magnetic field, it can be assumed that this is also valid for the toroidal component (Petrovay, 2004, private communication). This is why we applied an area-weighted method, since the sunspot area probably reflects the toroidal magnetic flux.

First, we confirmed the results of \citet{pulkkinenetal99} using the area-weighted mean latitude data. However, we found a longer period applying the discrete Fourier-transform and phase-dispersion method, respectively. Based on the $\chi^2$ test, the period of $P_\mathrm{PHD}=37\,900$ days with an amplitude of $1.04\degr$ turned out the best. We also studied the difference between the ensemble average area of sunspot groups for the two hemispheres, i.e.~ $(A_{30}/N_{30})_{\mathrm{North}}-(A_{30}/N_{30})_{\mathrm{South}}$, which also indicates that the dominance of the toroidal magnetic flux oscillates between the southern and northern hemispheres in a long time scale. For this dataset we a found somewhat longer period ($P=48\,500$ days). Although the deviation of the two periods is greater than the formal errors ($\pm100$ days), the real uncertainty is significantly larger due to the shortness of the observing window with respect to the length of the cycle and the large scatter. Thus, we do not think that the difference of the periods implies real physical effects. 

\citet{pulkkinenetal99} interpreted this latitudinal variation as a mixed parity mode in which a quadrupolar component is oscillating with this period. They examined the toroidal components of both the quadrupolar and the dipolar fields based on the paper of \citet{Brooke_etal98}. They claimed that multiplying the absolute value of the total toroidal field by the signed mean latitude of the activity belt, this product should oscillate about zero. As the difference between the ensemble average area of sunspot groups for the two hemispheres, i.e.~$(A_{30}/N_{30})_{\mathrm{North}}-(A_{30}/N_{30})_{\mathrm{South}}$ is related to the strength and pattern of the toroidal magnetic field, our results may indicate observational evidence of such an oscillation. We note that \citet{pipin99} also found an asymmetry of magnetic activity.

Beside studying the secular variation of the magnetic activity, we also investigated a further aspect of Gleissberg cycle. In Fig.~\ref{fig:cshfull_intime}-\ref{fig:avszfull_intime} it is visible that the sequential Schwabe cycles overlap each other at certain times. To characterize the rate of the overlapping we defined the daily latitudinal area-weighted extension of activity. We examined its mean and its standard deviation around the minima of the 11-year solar cycle. 
These quantities are increasing as the Gleissberg cycle tends to its maximum (see in Fig.~\ref{fig:avsz_mean_stdev}). While in the northern hemisphere this tendency is obvious, in the southern one it is less so, but the tendency is visible. Nevertheless, as our investigation is limited to one Gleissberg cycle, we cannot declare unambiguously that this phenomenon correlates with the Gleissberg cycle. 
\citet{Pelt_etal00} found a similar tendency in the distances in between nearby cycles \citep[see Fig.~11 in][]{Pelt_etal00}.
Hence, the continuous production of detailed, homogenous catalogues (e.g.~Debrecen Photoheliographic Data successor of the GPR) is crucial from this point of view. On the other hand, the interpretation of the variable amount of overlap beetween the Schwabe cycle is a challenge for future solar dynamo models.

\begin{acknowledgements}
We would like to thank K. Petrovay for fruitful discussions on the manuscript. This work was partly funded by the OTKA under grant no. T043741. B.~M. is grateful for support from the European Solar Magnetism Network, funded by the European Commission under contract HPRN-CT-2002-00313.
\end{acknowledgements}

\end{document}